\documentclass[11pt]{article}

\usepackage{ams}
\usepackage{amssymb}
\usepackage{graphicx}
\usepackage{caption}
\usepackage{pgf}
\usepackage{tikz}
\usepackage{comment}
\usepackage{amsthm}
\usepackage{verbatim}

\usetikzlibrary{arrows,automata}

\usepackage{latexsym}
\usepackage{amsfonts}
\usepackage[english]{babel}
\usepackage{xy}
\usepackage{fancyhdr}
\usepackage[a4paper,top=3cm,bottom=3.5cm,left=2.3cm,right=2.3cm,%
bindingoffset=5mm]{geometry}

\usepackage{setspace}

\newtheorem{proposition}{Proposition}
\newtheorem{remark}{Remark}

\begin{document}

\pagestyle{empty}

\centerline{\usefont{OT1}{phv}{b}{n}\selectfont
\fontsize{14pt}{10pt}\selectfont An ordinal measure of interrater absolute agreement}
\vspace{1cm}
\centerline{ Giuseppe Bove$^{*}$ , Pier Luigi Conti$^{**}$, Daniela Marella$^{*}$ 
\footnote{email: pierluigi.conti@uniroma1.it, giuseppe.bove@uniroma3.it,daniela.marella@uniroma3.it\\
Corresponding author: Giuseppe Bove - Dipartimento di Scienze della Formazione, Universit\`{a} ``Roma Tre'' - giuseppe.bove@uniroma3.it}}
\vspace {.5cm}
\centerline{{\it$^{*}$Dipartimento di Scienze della Formazione, Universit\`{a} ``Roma Tre''}}
\centerline{{\it $^{**}$Dipartimento di Scienze Statistiche, Universit\`{a} ``La Sapienza''}}

\bigskip

\begin{abstract} 
A measure of interrater absolute agreement for ordinal scales is proposed capitalizing on the dispersion index for ordinal variables proposed by Giuseppe Leti. The procedure allows to avoid the problem of restriction of variance that sometimes affect traditional measures of interrater agreement in different fields of application. An unbiased estimator of the proposed measure is introduced and its sampling properties are investigated. 
In order to construct confidence intervals for interrater absolute agreement both asymptotic results and bootstrapping methods are used and their performance is evaluated. Simulated data are employed to demonstrate the accuracy and
practical utility of the new procedure for assessing agreement.
Finally, an application to a real case is
provided.
\end{abstract}


\section{Introduction}
\label{sec:1}
Ordinal rating scales are frequently developed in study designs where several ‘raters’ (or ‘judges’) evaluate a group of ‘targets’. For instance, in language studies new rating scales before their routine application are tested out by a group of raters, who assess the language proficiency of a corpus of argumentative (written or oral) texts produced by a group of writers. Similar situations can be found in organizational, educational, biomedical, social, and behavioural research areas, where raters can be counsellors, teachers, clinicians, evaluators, or consumers and targets can be organization members, students, patients, subjects, or objects. 
When each rater evaluates targets, the raters provide comparable categorizations of the targets. The more the raters categorizations coincide, the more the rating scale can be used with confidence without worrying about which raters produced those categorizations. Hence, the main interest here consists in analysing the extent that raters assign the same (or very similar) values on the rating scale (interrater absolute agreement), that is to establish to what extent raters evaluations are close to an equality relationship (\emph{e.g.}, in the case of only two raters, if the two sets of ratings are represented by $x$ and $y$ the relation of interest is $x = y$).  
Measures of interrater absolute agreement, as Cohen's Kappa (and extensions to take into account three or more raters, e.g., \cite{von:2005}) and intraclass correlations (\cite{shrout:1979}; \cite{McGraw:1996}) are usually applied when dealing with rating performed by ordinal scales. A first problem of these procedures is that they are not originally defined for ordinal scales, and so they have to be adapted.
For instance,  the application of indices based on Cohen's Kappa need to assign numerical values to the ordinal level of the scale; intraclass correlation indices are based on ANOVA for repeated measures approach for interval data. Another limitation of the above mentioned measures is that they are affected by the restriction of variance problem (e.g., \cite{Lebreton:2003}), that consists in an attenuation of estimates of rating similarity caused by an artefact reduction of the between-subjects variance in ratings. For instance, this happens in language studies when the same task is defined for native (L1) and non-native (L2) writers, and the analysis compare rater agreement in the two groups separately. Even in the presence of a very good absolute agreement, Cohen's Kappa coefficient and intraclass correlations can take low values, especially for L1 group, because the range of ratings provided by the raters are concentrated on one or two very high levels of the scale (a range restriction that determines a between-target variance restriction). 

In order to overcome the restriction of variance problem, measure for absolute agreement (or consensus) have been proposed, see \cite{Lebreton:2008} for a review. The main underlying idea is to measure the within-target variance of ratings (i.e., the between-rater variance) separately for each target, and summarize the results in a final average index (usually normalized in the interval $[0-1]$). In this approach, the influence of the low level of the between-target variance is removed by separate analysis of the ratings of each target. One of the most popular index in this group was proposed by (\cite{demaree:1984},\cite{demaree:1993}). For a scale $X$ it can be expressed as
\begin{eqnarray}
r_{WG}=1-\frac{s^{2}_{X}}{\sigma^2_{E}}
\end{eqnarray}
where $s^{2}_{X}$ is the observed between-rater variance of the ratings and $\sigma^2_{E}$ is the between-rater variance obtained from a theoretical null distribution representing a complete lack of agreement among raters. Roughly speaking, the null distribution conceptually represents no agreement,
which means that to calculate $r_{WG}$, one makes a direct comparison
between the observed variance in raters' ratings with
the variance one would expect if there was no agreement among
raters. Higher numbers indicate a greater agreement. 

For raters in perfect agreement we have $s^{2}_{X}=0$, with a corresponding value $r_{WG}=1$. In applications,  $r_{WG}$ values greater than $0.7$ (possibly $0.8$) are considered associated with high level of interrater absolute agreement (see \cite{Lebreton:2008}, p. 836 table 3). Often
researchers define the no agreement, or the null distribution, in
terms of a uniform distribution.  When the null distribution is assumed as uniform, the equation for the corresponding variance $\sigma^2_{EU}$ is 
\begin{eqnarray}
\sigma^2_{E}=\sigma^2_{EU}=\frac{K^2-1}{12}
\end{eqnarray}
where $K$ refers to the total number of levels of the scale $X$.

The index $r_{WG}$ and  other indices reviewed in \cite{Lebreton:2008} (\emph{e.g.}, standard and average deviation indices) allow to avoid the problem of variance restriction, but as traditional measures of interrater agreement they are defined only for interval data. Besides, depending on the choice of the null distribution, negative values could be obtained.   
For these reasons, in this contribution we propose a new procedure to measure absolute agreement for ordinal rating scales by using the dispersion index proposed by \cite{leti:1983} (pp. 290-297) for ordinal variables. In this way, we take into consideration the ordinal level of the measurement scales. The new measure is not affected by restriction of variance problems and does not depend on the choice of a particular null distribution. 
In this paper we assume a two-way random sampling design, where the sampling design involves a sample of raters as well as a sample of targets, all of which are rated by each sampled rater.

The paper is organized as follows. In Section \ref{sez_1} the dispersion index proposed by \cite{leti:1983} (pp. 290-297) for ordinal variables is introduced and its sampling properties are analyzed in Section \ref{sez_2}. Section \ref{sez_3} contains results of a simulation experiment used to illustrate the theoretical results. With this regard, confidence intervals for the proposed interrater agreement index are constructed using both the asymptotic results described in Section \ref{sez_2} (Proposition 4) and bootstrapping procedures. Finally, in Section \ref{sez_4} an application to real data is performed.

\section{Leti index as a measure of interrater absolute agreement for ordinal scales}
\label{sez_1}
The dispersion of an ordinal categorical variable can be measured by the index proposed in \cite{leti:1983} (pp. 290-297), which is given by
\begin{eqnarray}
\label{base}
D=2\sum_{k=1}^{K-1}F_{k}(1-F_{k})
\end{eqnarray}
where $K$ is the number of categories of the variable $X$ and  $F_k$ is the cumulative proportion associated to category $k$, for $k=1,\dots,K$. Index $(\ref{base})$ is nonnegative and it is easy to prove that  $D=0$ if and only if all observed categories are equal (absence of dispersion). The maximum value of the index ($D_{max}$) is obtained when all observations are concentrated in the two extreme categories of the variable (maximum dispersion), and it is
 \begin{eqnarray}
D_{max}=\frac{K-1}{2} 
\end{eqnarray}
as $N$ is even,
\begin{eqnarray}
D_{max}=\frac{K-1}{2} \left(1-\frac{1}{N^2}\right)
\end{eqnarray}
as $N$ is odd, $N$ being the total number of observations. For $N$ moderately large, the maximum of the index can be assumed equal to $(K-1)/2$. Hence, it is possible to define a measure  of dispersion normalized in the interval $[0,1]$ given by 
\begin{eqnarray}
\label{index_d}
d=\frac{D}{D_{max}}=\frac{2}{K-1}D.
\end{eqnarray}
Two advantages of this proposal respect to measures of absolute agreement like $r_{WG}$ reported below are: 
\begin{enumerate}
\item[(i)] $d$ does not depend by the formulation of a null distribution for normalization;
\item[(ii)] $d$ can never be out of the range $[0,1]$. 
\end{enumerate}
It is interesting to notice that $D$ has properties of within and between dispersion decomposition analogous to the well-known variance decomposition \cite{grilli:2002}.

\section{Sampling Properties of $d$ index}
\label{sez_2}

A sample of $n_R$ raters and a sample of $n_T$ targets are drawn by simple random
sampling without replacement from a finite population of targets and raters, respectively. Let us denote with $X_{ij}$ the score given by the $j$th rater to the $i$th target on a $K$-point scale, for $i=1,\dots,n_T$ and $j=1,\dots, n_R$. Formally, $X_{ij}$s are independent categorical random variables having $K$ categories with $p_{k}^{(ij)}=P(X_{ij}=k)$, for $i=1,\dots,n_T$, $j=1,\dots, n_R$ and $k=1,\dots,K$.
In the sequel we assume that both the targets and the raters are homogeneus (\emph{targets-raters homogeneity assumption}), which implies that the probability $p_{k}^{(ij)}=p_{k}$ does not depend on rater $j$ or target $i$, for $i=1,\dots,n_T$, $j=1,\dots, n_R$, $k=1,\dots,K$. 

As a consequence of \emph{homogeneity assumptions}, the  variables $X_{ij}$ are independent and identically distributed (\emph{i.i.d.}). As previously stressed, the dispersion of an ordinal categorical variable can be measured by the index (\ref{base}).
With regard to $i$th target, let us denote
with $\widehat{F}_{k}^{(i)}$ the empirical cumulative distribution function defined as
\begin{eqnarray}
\widehat{F}_{k}^{(i)}=\frac{1}{n_R}\sum_{j=1}^{n_R}I_{(X_{ij}\leq k)}
\end{eqnarray}
where the numerator represents the number of raters giving score less than or equal to $k$ to the $i$th target. It is known that
$E(\widehat{F}_{k}^{(i)})=F_{k}^{(i)}=F_{k}$, where the last equality comes from the \emph{targets homogeneity assumptions}.
Furthermore, $V(\widehat{F}_{k}^{(i)})=F_{k}(1-F_{k})$ and 
$Cov(\widehat{F}_{k}^{(i)},\widehat{F}_{l}^{(i)})=min(F_{k},F_{l})-F_{k}F_{l}$.

In order to estimate $(\ref{base})$, for each target $i$ the following estimator can be defined
\begin{eqnarray}
\label{leti_1}
\widehat{D}_{i}=2\sum_{k=1}^{K-1}\widehat{F}_{k}^{(i)}(1-\widehat{F}_{k}^{(i)}).
\end{eqnarray}

As stressed in \cite{piccarretta:2001}, (\ref{leti_1}) can be alternatively expressed as 
\begin{eqnarray}
\label{est_leti_1}
\widehat{D}_{i}&=&\sum_{k=1}^{K}\sum_{l=1}^{K}|k-l|\widehat{p}_{k}^{(i)}\widehat{p}_{l}^{(i)}\nonumber\\
&=&\frac{1}{n_{R}^2}\sum_{j=1}^{n_R}\sum_{j^{'}=1}^{n_R}|X_{ij}-X_{ij^{'}}|
\end{eqnarray}
where 
\begin{eqnarray}
\label{p_est}
\widehat{p}_{k}^{(i)}=\frac{1}{n_R}\sum_{j=1}^{n_R}I_{(X_{ij}= k)}
\end{eqnarray}
is an unbiased estimator of $p_k$. 

\begin{proposition}
\label{prop_1}
The random variable (r.v) $n_R(\widehat{p}_{1},\dots, \widehat{p}_{K})^{'}$, with 
$\widehat{p}_{k}=\sum_{i=1}^{n_{T}}\widehat{p}^{(i)}_{k}/n_{T}$ for $k=1,\dots,K$, 
 follows a multinomial distribution with parameters $n_Rn_T$ and $(p_{1},\dots, p_{K})$.
\end{proposition}

The expression (\ref{est_leti_1}) allows to compute easily the expectation and the variance of estimator $(\ref{leti_1})$ as shown in Proposition \ref{prop_2}, see \cite{lomnicki:1952} for details.
\begin{proposition}
\label{prop_2}
The estimator $\widehat{D}_{i}$ has expectation
\begin{eqnarray}
\label{expectation}
E(\widehat{D}_{i})=\left( 1-\frac{1}{n_R}\right)D
\end{eqnarray}
and variance  given by
\begin{eqnarray}
\label{variance}
Var(\widehat{D}_{i})&=&\left(\frac{1}{n_R^2}-\frac{1}{n_R^3}\right)(4\sigma^2+4(n_R-2)J-2(2n_R-3)D^{2})=V
\end{eqnarray}
where 
\begin{eqnarray}
\sigma^2&=&Var(X_{ij})=\sum_{k=1}^{K}k^2p_{k}-\left(\sum_{k=1}^{K}kp_{k}\right)^{2}\\
J&=&\sum_{k=1}^{K}\sum_{h=1}^{K}\sum_{l=1}^{K}|k-h||k-l|p_{k}p_{h}p_{l}.\end{eqnarray}
 \end{proposition}
\begin{proof}
Both (\ref{expectation}) and (\ref{variance}) come from the results in \cite{lomnicki:1952}. With regard to (\ref{expectation}), we have
\begin{eqnarray}
E(\widehat{D}_{i})&=&E\left(\frac{1}{n_{R}^2}\sum_{j=1}^{n_R}\sum_{j^{'}=1}^{n_R}|X_{ij}-X_{ij^{'}}|\right)\nonumber\\ &=&
\frac{n_R(n_R-1)}{n_R^2}E\left(\frac{1}{n_R(n_R-1)}\sum_{j=1}^{n_R}\sum_{j^{'}=1}^{n_R}|X_{ij}-X_{ij^{'}}|\right)\nonumber\\ &=&
\frac{n_R(n_R-1)}{n_R^2}2\sum_{k=1}^{K-1}F_k(1-F_k)\nonumber\\ &=&
\left(\frac{n_R-1}{n_R}\right)D
\end{eqnarray}
for the variance (\ref{variance}) we obtain
\begin{eqnarray}
Var(\widehat{D}_{i})&=&Var\left(\frac{1}{n_R^2}\sum_{j=1}^{n_R}\sum_{j^{'}=1}^{n_R}|X_{ij}-X_{ij^{'}}|\right)\nonumber\\ &=&
\left(\frac{n_R-1}{n_R}\right)^2Var\left(\frac{1}{n_R(n_R-1)}\sum_{j=1}^{n_R}\sum_{j^{'}=1}^{n_R}|X_{ij}-X_{ij^{'}}|\right)\nonumber\\&=&
\left(\frac{n_R-1}{n_R}\right)^2\frac{1}{n_R(n_R-1)}(4\sigma^2+4(n_R-2)J-2(2n_R-3)D^{2})\nonumber\\ &=&
\left(\frac{1}{n_R^2}-\frac{1}{n_R^3}\right)(4\sigma^2+4(n_R-2)J-2(2n_R-3)D^{2}).
\end{eqnarray}
\end{proof}
\begin{remark}
For $n_R$ sufficiently large, we have
\begin{eqnarray}
Var(\widehat{D}_{i})\approx \frac{4(J-D^{2})}{n_R}.
\end{eqnarray}
\end{remark}

As an  estimator of $d$ index (\ref{index_d}) we consider
\begin{eqnarray}
\widehat{d}=\frac{\widehat{\overline{D}}}{D_{max}}=\frac{1}{D_{max}}\left(\frac{1}{n_T}\sum_{i=1}^{n_T}\widehat{D}_{i}\right).
\end{eqnarray}

where $\widehat{\overline{D}}$ is an estimator of $D$ obtained averaging the $n_T$ estimates $\widehat{D}_{1},\dots, \widehat{D}_{n_T}$.


In Proposition \ref{prop_3} both the sampling properties and the asymptotic distribution of
$\widehat{d}$ are analyzed for large $n_T$ and moderate $n_R$.
\begin{proposition}
\label{prop_3} The estimator $\widehat{d}$ has expectation
\begin{equation}
\label{biasD}
E(\widehat{d})=\left(\frac{n_R-1}{n_R}\right)d
\end{equation}
and variance 
\begin{eqnarray}
\label{vard}
V_{d}=\left(\frac{1}{D_{max}}\right)^2 \frac{V}{n_T}
\end{eqnarray}
where $V$ is given in $(\ref{variance})$.
Furthermore, since $\widehat{D}_{1},\dots,\widehat{D}_{n}$ are \emph{i.i.d.}, for the central limit theorem, as $n_T$ goes to infinity the random variable $\widehat{d}$ tends to a standard normal distribution with mean and variance given by (\ref{biasD}) and (\ref{vard}), respectively,
\end{proposition}
In Proposition \ref{prop_4} an unbiased estimator of $d$ is proposed and its asymptotic distribution  is evaluated.

\begin{proposition}
\label{prop_4}
From (\ref{biasD}), an unbiased estimator of $d$ can be defined as follows
\begin{eqnarray}
\label{D_unb}
\widehat{d^{*}}=\frac{n_R}{n_R-1}\widehat{d}.
\end{eqnarray}
As a consequence of Proposition (\ref{prop_3}), the distribution of  $\widehat{d}^{*}$  is approximately normal with mean $d$ and variance
\begin{eqnarray}
\label{var_star}
V_{d^{*}}=\left(\frac{n_R}{n_R-1}\right)^2\left(\frac{1}{D_{max}}\right)^2 \frac{V}{n_T}.  
\end{eqnarray}

\end{proposition}

The proof of Proposition \ref{prop_4} follows from Proposition \ref{prop_3}.
The above results are useful to construct point and interval estimates of  $d$. They are also useful 
for testing both the statistical significance of the index (that is the null hypotheses $H_{0}: d=0$) and null hypothesis such as  
$H_{0}: d \leq d_{0}$, where $d_{0}$ be a real number in $[0,1]$. Consider the
hypothesis problem 

\begin{equation}
\label{hypothesis}
\bigg \{
\begin{array}{rl}
 H_0:& d \leq d_{0}  \\
 H_1: &d >d_{0} \\
\end{array}
\end{equation}
As a consequence of
Proposition \ref{prop_4}, a test with an asymptotic significance level $\alpha$ consists in accepting $H_0$ whenever
\begin{eqnarray}
\widehat{d}^{*} \leq d_{0}+ z_{\alpha}\sqrt{\widehat{V}_{d^{*}}}
\end{eqnarray}
where $z_\alpha$ is the $\alpha-th$ quantile of the standard normal distribution
and $\widehat{V}_{d^{*}}$ is an estimate of variance (\ref{var_star}).

\section{Simulation Study}
\label{sez_3}
In this section, a simulation study to compare the performance of different confidence intervals
for index $d$ is performed. We focus our efforts on developing methods for constructing confidence intervals
for the index $d$ because confidence
intervals indicate the range within which the population
parameter $d$ (the interrater agreement in the population) is likely to fall, as well as
precision of this estimate (i.e., the size of the range). 

A finite population of size $N_T=150$ targets and $N_R=28$ raters was generated from a multinomial model with parameters $N_R=28$ and probabilities $(p_1,p_2,p_3,p_4,p_5)=(0.1,0.2,0.35,0.25,0.1)$.
Then, the finite population consists in a matrix $P$ of size $N_T \times N_R$.
The value of $d$ index (\ref{index_d}) is $0.61$.

From the population, $S=1000$ samples were drawn according to a simple random sampling without replacement on the basis of the following two-step procedure. First of all, a simple random sample of size $n_R=7$ from the $N_R=28$ raters has been selected. This  is equivalent to select a simple random sampling without replacement of columns in the finite population matrix $P$, the result is a matrix $P_R$ of size $N_T \times n_R$. Secondly, a simple random sampling of size $n_T=50$ from $N_T=150$ targets has been drawn. 
This means to draw a simple random sampling of $n_T=50$ rows from $P_R$.

In order to construct confidence intervals
for the index $d$, both the asymptotic result in Proposition \ref{prop_4} and bootstrapping procedures are used. The bootstrap methods are described in points (2)-(4) below, where we assume that $B = 1000$ bootstrap samples are drawn from each initial
sample $s$.
Formally, confidence intervals for $d$ of level $1-\alpha=0.95$ have been constructed using the following methods:
\begin{enumerate}
\item[(1)] \emph{Normal approximation}. For the initial sample $s$ (for $s=1,\dots,S$), the confidence interval $[L^{s}_{Norm},U^{s}_{Norm}]$ based on the asymptotic normal approximation is given by
\begin{eqnarray}
L^{s}_{Norm}=\widehat{d}^{*}-z_{1-\alpha/2}\sqrt{\widehat{V}_{d^{*}}}; \quad
U^{s}_{Norm}=\widehat{d}^{*}-z_{\alpha/2}\sqrt{\widehat{V}_{d^{*}}}
\end{eqnarray}
where $\widehat{d}^{*}$ and $\widehat{V}_{d^{*}}$ are the estimates of $d$ and $V_{d^{*}}$, respectively.
\item[(2)] \emph{Percentile method}. For the initial sample $s$ (for $s=1,\dots,S$), the confidence interval $[L^{s}_{Perc},U^{s}_{Perc}]$ is obtained 
by taking $\alpha/2$ and $1-\alpha/2$ quantiles of the $B$ bootstrap samples.
Formally
\begin{eqnarray}
L^{s}_{Perc}=Q_{\alpha/2}; \quad
U^{s}_{Perc}=Q_{1-\alpha/2}
\end{eqnarray}
\item[(3)]  \emph{Bootstrap-t interval}. For the initial sample $s$ (for $s=1,\dots,S$), the confidence interval $[L^{s}_{T},U^{s}_{T}]$ is computed as follows
\begin{eqnarray}
L^{s}_{T-int}=\widehat{d^{*}}-t_{1-\alpha/2}\sqrt{\widehat{V}_{d^{*}}}; \quad
U^{s}_{T-int}=\widehat{d^{*}}-t_{\alpha/2}\sqrt{\widehat{V}_{d^{*}}}
\end{eqnarray}
where $t_{\alpha}$ is the $\alpha$th percentile of the distribution of $z^{*}_{b}$ (for $b=1,\dots,B$)  with
\begin{eqnarray}
\label{z_int1}
z^{*}_{b}=\frac{\widehat{d^{*}_{b}}-\widehat{d^{*}}}{\widehat{se}^{*}_{b}}.
\end{eqnarray}
In (\ref{z_int1}) $\widehat{d^{*}_{b}}$ is the estimate of $d^{*}$ based on the  $b$th bootstrap sample and $\widehat{se}^{*}_{b}$ is the standard error based on the data in the $b$th bootstrap sample. 
\item [(4)] \emph{Pivotal method}. For the initial sample $s$ (for $s=1,\dots,S$), the confidence interval $[L^{s}_{Pivot},U^{s}_{Pivot}]$ is computed as follows
\begin{eqnarray}
L^{s}_{Pivot}=2\widehat{d^{*}}-Q_{1-\alpha/2}; \quad
U^{s}_{Pivot}=2\widehat{d^{*}}-Q_{\alpha/2}
\end{eqnarray}
where $Q_{\alpha/2}$ and $Q_{1-\alpha/2}$ are the $\alpha/2$ and $1-\alpha/2$ quantiles of the $B$ bootstrap estimates 
$\widehat{d^{*}_{b}}$, for $b=1,\dots,B$. 
\end{enumerate} 

As far as the methods described in steps (2)-(4) are concerned, from each of the $S=1000$
initial samples, the $B=1000$ bootstrap samples were selected according to the following methods:
\begin{enumerate}
 \item[1] \emph{Nonparametric bootstrap}. 
From each initial sample $s$, the $b$th bootstrap sample is selected as follows: (i) a simple random sample with replacement of $r=7$ raters has been selected from the original sample of raters; (ii) a simple random sampling with replacement of $n=50$ writers has been drawn from the original sample of writers.
\item[2] \emph{Parametric bootstrap} From each initial sample $s$, the $b$th bootstrap sample is generated according the multinomial model specified in Proposition \ref{prop_1}.
\item[3] \emph{Pseudo-Nonparametric bootstrap}. The nonparametric bootstrap described in point (1), is based on the assumption that the data are \emph{i.i.d.}, see \cite{efron:1979}.
Since survey data are not necessarily \emph{i.i.d.}, many
bootstrap resampling methods have been proposed in the context of survey sampling. These methods are obtained after making some
modifications to the classical \emph{i.i.d.} bootstrap in order to adapt it for survey data.  For a review of bootstrap methods in the context of survey data, see \cite{mashreghi:2016}.
The class of pseudo-population bootstrap methods 
consists in creating
a pseudo-population by repeating the units of the initial sample and drawing from such a pseudo-population 
bootstrap samples with the same design as the initial one.
In order to illustrate how a pseudo-population is constructed, let us assume that a simple random sample without replacement has been selected from a finite population of size $N$.
A pseudo-population of size $N$ can be created by repeating the selected sample, $N/n$ times.
This method, was first introduced by \cite{gross:1980}. In practice $N/n$ is rarely an integer, in this case a method to build a pseudo-population of size $N$ was proposed by \cite{booth:1994}.
In this method,  a pseudo-population is first constructed by replicating $k=\lfloor{N/n\rfloor}$ times each unit of
the original sample $s$. Then, the pseudo-population is completed by taking a
simple random sample of size $N-nk$ without replacement from $s$. 
Taking into account the two-way sampling design of both targets and raters, the pseudo-population has been generated according the following two step procedure:
\begin{enumerate}
\item[Step 1]  the ratings of $N_R=28$ raters have been reconstruted replicating the columns of the original sample $s$, $k_R=N_R/n_R=28/7=4$ times. As a consequence, this first step generates a sample $s_R$ of size $n_T=50$ and $n_R=N_R=28$;
\item[Step 2] the points of  $N_T=150$ targets have been reconstruted replicating the rows of the sample $s_R$ obtained in Step 1, $k_T=N_T/n_T=150/50=3$ time.  
\end{enumerate}
\end{enumerate}


The accuracy of confidence intervals has been evaluated
by the following indicators.
\begin{enumerate}
\item[(1)] Estimated coverage probability, in per cent, for the interval
\begin{eqnarray}
ECP=\frac{100}{S}\sum_{s=1}^{S}I(L_{t}^{s} \leq d \leq U_{t}^{s})
\end{eqnarray}
\item[(2)] Estimated left-tail and right-tail errors (lower and upper error rates) in per cent
\begin{eqnarray}
LE&=&\frac{100}{S}\sum_{s=1}^{S}I(L_{t}^{s} >d)\\
RE&=&\frac{100}{S}\sum_{s=1}^{S}I(U_{t}^{s} <d),
\end{eqnarray}
\item[(3)] Estimated average length (AL) of all 1000 simulated intervals given by
\begin{eqnarray}
AL=\sum_{s=1}^{S}\frac{U_{t}^{s}-L_{t}^{s}}{S}
\end{eqnarray}
\end{enumerate}
where $I(a)=1$ if $a$ is true and $I(a)=0$ elsewhere, and $t=Norm,T-int,Perc,Pivot$. 


\subsection{Simulation results}
Tables \ref{table1} presents the outcomes achieved in the simulation study. More specifically, the estimated
coverage probabilities of $95\%$ confidence intervals (CP), the estimated left-tail (LE) and right-tail (RE)
errors (nominal values is $2.5\%$ for both) and the average length (AL) for the index $d$, when ($n_R=7,n_T=50$),  are reported. 
The $d$ value  is equal to $0.61$.

{\small
\begin{table}[!ht]
\caption{\emph{Performance of different confidence intervals for d when $n_R=7$, $d=0.61$}} \label{table1}
\begin{center}
\begin{tabular}{|c|c|c|c|c|}
\hline
\hline
\multicolumn{2}{|c|}{} & \multicolumn{3}{|c|}{$n_R=7$}\\
\hline
{Method} & {Indicators} & { $Nonparametric$ }   &  { $Parametric$} & {$Pseudo-Nonparametric$}\\
\cline{1-2}
\hline
\hline
{\it Normal}&CP	&99.4  & 	99.4& 99.4\\
&LE& 0.6 &0.6&0.6 \\
&RE	& 0&0&0\\
&AL& 0.16&0.16&0.16\\
\hline
\hline
{\it T-int}&CP& 26.2& 72.4& 28.8\\
&LE&  73.8& 26.2&71.2\\
&RE	& 0&1.4&0\\
&AL& 0.18&0.08&0.15\\
\hline
\hline
{\it  Perc}&CP	&92.8& 91.2&92.8\\
&LE&  0& 8.8&0\\
&RE	&  7.2& 0&7.2\\
&AL& 0.23 & 0.10 & 0.18\\
\hline 
\hline
{\it Pivot}& CP	&27& 79.2&30\\
&LE&  73& 19.6&70\\
&RE	& 0 &1.2&0\\
&AL& 0.23&0.10&0.18\\
\hline
\end{tabular}
\end{center}
\end{table}
}

\begin{center}
\begin{figure}[!ht]
 \centering 
\includegraphics[scale=.85]{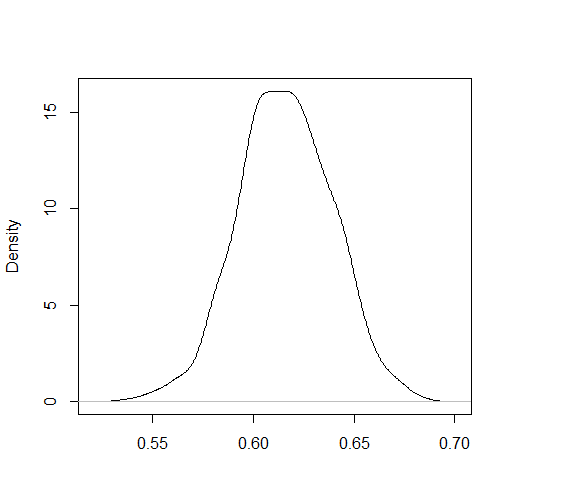}
\caption{Kernel density estimate of $d$ index from the 1000 original samples.}
\label{figure0}
\end{figure}
\end{center}

As reported in Table \ref{table1}, the confidence intervals obtained with the normal approximation perform very well.
Coverage probabilities are larger than $95\%$ nominal value ($99.4\%$) with an average length of $0.16$.
Furthermore, the normal confidence intervals construction is simple, as it does not require resampling from the
initial sample. Figure \ref{figure0} shows the kernel density of the $d$ index estimated from the 1000 original samples. 
The bandwidth selection rule is as proposed by \cite{sheather:1991}. 

The percentile method has a good performance with coverage probability larger than $91\%$. 
The worst methods are the $Pivot$ and $T-int$ methods. The lower and upper error
rates, giving us an idea of how skewed the distribution of the $d$ estimator is, are not well
balanced.
With regard to the methods used to generate the bootstrat samples, the \emph{parametric} approach performance  is strictly related to the estimation of the multinomial probabilities.
As previously stressed,  each row in the inital sample $s$ provides an estimate of $(p_1,p_2,p_3,p_4,p_5)$ and the mean of such estimates defines the estimated probabilities $(\widehat{p}_1,\widehat{p}_2,\widehat{p}_3,\widehat{p}_4,\widehat{p}_5)$ of the multinomial distribution used to generate the bootstrap samples as specified in Proposition \ref{prop_1}. In Table \ref{table2}, 
the minimum, the maximum, the mean and the standard deviation of the distribution of $\widehat{p}_k$ (for $k=1,2,3,4,5$) estimated from the original 1000 samples are reported. 
{\small
\begin{table}[!ht]
\caption{\emph{Descriptive statistics of $\widehat{p}_k$ distribution, for $k=1,2,3,4,5$ and $d=0.61$}} \label{table2}
\begin{center}
\begin{tabular}{|c|c|c|c|c|c|}
\hline
{\it Parameter} & {\it True value} & {\it Min }   &  {\it Max} & {\it Mean} & {\it Sd}\\
\hline
\hline
$p_1$ &0.10& 0.05  & 0.15& 0.10 &0.01\\
$p_2$ &0.20&0.14  & 0.25& 0.19 &0.02\\
$p_3$ &0.35&0.29  & 0.41& 0.35 &0.02\\
$p_4$ &0.25&0.20  & 0.33& 0.26 &0.02\\
$p_5$ &0.10&0.06  & 0.16& 0.10 &0.02\\
\hline
\end{tabular}
\end{center}
\end{table}
}



As  Table \ref{table1} shows, the \emph{pseudo-nonparametric} approach taking into account the sample selection effects has a slightly better performance than the \emph{nonparametric} approach both in terms of coverage probabilities and  average lengths for all methods ($T-int$, $Perc$, $Pivot$).

Finally, note that in the \emph{nonparametric} approach the resampling with replacement from $n_R=7$ raters generates a replication of columns of the bootstrap sample introducing a false agreement between raters and as a consequence an underestimation of $d$.
This fact is showed in Table \ref{tablebias} where the mean of the $d$ estimates over both  the $1000$ original samples $s$ and over the bootstrap replications $b$ are reported. 

Such means have been computed both for the original population with $d=0.61$ and for a population with $d=0.41$, showing as the magnitude of bias depends also on the original agreement degree between raters. 
That is, the higher the raters agreement (low values of $d$), the smaller the bias in the $d$ estimator introduced by the resampling with replacement.
Clearly, such a bias is also present in the \emph{pseudo-nonparametric} approach but with a smaller magnitude, thank to the construction of the pseudo-population that mitigates such a phenomenon.  As table \ref{tablebias} shows, the \emph{parametric approach} produces null bias estimates.
{\small
\begin{table}[!ht]
\caption{\emph{The mean of $\widehat{d}$ over the initial samples $s$ and over the bootstrap replications $b$.}} \label{tablebias}
\begin{center}
\begin{tabular}{|c|c|c|}
\hline
{\it Approach} & {\it mean of $\widehat{d}^*$  (d=0.61)} & {\it mean of $\widehat{d}^*$  (d=0.41)}  \\
\hline
\hline
{\it Nonparametric} &0.53&0.36\\
{\it Parametric} &0.61&0.41\\
{\it Pseudo-nonparametric} &0.55&0.37 \\
\hline
\end{tabular}
\end{center}
\end{table}
}

The simulation in Table \ref{table1} has been repeated for a populaiton with $d=0.41$. The results are reported in Table \ref{table1pic}.

{\small
\begin{table}[!ht]
\caption{\emph{Performance of different confidence intervals for d when $n_R=7$, $d=0.41$}} \label{table1pic}
\begin{center}
\begin{tabular}{|c|c|c|c|c|}
\hline
\hline
\multicolumn{2}{|c|}{} & \multicolumn{3}{|c|}{$n_R=7$}\\
\hline
{Method} & {Indicators} & { $Nonparametric$ }   &  { $Parametric$} & {$Pseudo-Nonparametric$}\\
\cline{1-2}
\hline
\hline
{\it Normal}&CP	&98.2  & 	98.2& 98.2\\
&LE& 1.8 &1.8&1.8 \\
&RE	& 0&0&0\\
&AL& 0.13&0.13&0.13\\
\hline
\hline
{\it T-int}&CP& 60.2& 83.2& 61.2\\
&LE&  39.8& 14.8&38.8\\
&RE	& 0&2&0\\
&AL& 0.18&0.10&0.14\\
\hline
\hline
{\it  Perc}&CP	&93.2& 93.8&93.2\\
&LE&  0& 5.8&0\\
&RE	&  6.8& 0.4&6.3\\
&AL& 0.19 & 0.10 & 0.15\\
\hline 
\hline
{\it Pivot}& CP	&64.8& 84.6&65.4\\
&LE&  35.2& 12.6&34.6\\
&RE	& 0 &2.8&0\\
&AL& 0.19&0.10&0.15\\
\hline
\end{tabular}
\end{center}
\end{table}
}

In conclusion, the most competitive method in terms of performance and computational
time seem to be the normal. Among the bootstrapping procedures the percentile method in the \emph{parametric} approach seems to perform better.

\section{An application on real data: the assessment of language proficiency}
\label{sez_4}
The aim of this section is to apply the methodology illustrated in the previous sections on an empirical data set,
we have analysed ratings obtained in a research conducted at Roma Tre University
(see \cite{nuzzo:2018}, for a detailed description). The main aim
of the study was to investigate the applicability of a six-point Likert scale for
functional adequacy (an aspect of language proficiency) developed by \cite{kuiken:2017}
 to texts produced by native and non-native writers, and to
different task types (narrative, instruction, and decision-making tasks). The
scale comprises four subscales, corresponding to the four dimensions of functional
adequacy identified by the authors of the scale: content, task requirements,
comprehensibility, coherence and cohesion (the reader is referred to
\cite{kuiken:2017} for a detailed presentation of scales and descriptors).
20 native speakers of Italian (L1) and 20 non-native speakers of Italian
(L2) participated in the study as writers. All the texts produced by L1 and L2
writers (120 texts in total for the three tasks) were assessed by 7 native speakers
of Italian on the Kuiken and Vedder’s six-point Likert scale. The raters did
not have any specific experience in judging written texts, and can therefore be
categorized as being non-expert. For our purposes, we have selected ratings
concerning only the narrative task and the subscale comprehensibility.
Just to give a general idea of the subscale, definitions of levels 1 and 6 are reported in the following:
\begin{enumerate}
\item[Level 1:] The text is not at all comprehensible. Ideas and purposes are unclearly stated and the efforts of the reader to understand the text are ineffective.
\item[Level 6:] The text is very easily comprehensible and highly readable. The ideas and the purpose are clearly stated.
\end{enumerate}

The results of the interrater agreement analysis for the subscale are summarized
in Table \ref{tab_1}, where the intraclass correlation $ICC(A,1)$ and the average values of $r_{WG}$, as defined in \cite{Lebreton:2008}, the coefficient of variation $CV$, $\widehat{d}$ and $\widehat{d}^{*}$
are shown for L1, L2 and total groups.
The intraclass correlation  $ICC(A,1)$ provides a low-moderate level of agreement for the total group
($0.67$). The results for the average values of
$CV$ ($12.16\%$), $\widehat{d}$ ($0.22$) and $\widehat{d}^{*}$ ($0.25$) seem in accord with $ICC(A, 1)$, while the average
value of $r_{WG}$ (0.87), highlights
a higher level of agreement. As it was observed in \cite{bove_2018}, when the analysis focuses separately on the two
subgroups of L1 and L2 students, results 
regarding the L1 group deserve
particular attention. Interrater agreement measured by intraclass correlation
is very low in the L1 group ($ICC(A, 1) = 0.14$). Analysing the dispersion
of the ratings given to this subgroup, it comes out that most of the raters used
almost exclusively levels 5 and 6 of the scale. Such a range restriction caused the very low value of the intraclass
correlation, despite the substantial agreement among the raters that scored all
the L1 texts in the same high levels. This problem does not regard the results
for the other three indices of Table \ref{tab_1} ($r_{WG} = 0.90$; $CV = 8.12\%$; $\widehat{d} = 0.17$; $\widehat{d}^{*}= 0.19$)
that show a very good level of absolute agreement.
Finally, the standard deviation of $\widehat{d}^{*}$ computed on the basis of formula (\ref{var_star}) is equal to $0.05$. 
As a consequence, the $(1-\alpha)=0.95$ confidence interval using the normal approximation for the total group is $[0.15, 0.35]$ and the error is at most $0.10$.

{\small
\begin{table}[!ht]
\caption{\emph{$ICC(A,1)$ and average of $r_{WG}$, $CV$, $\widehat{d}$ and $\widehat{d}^{*}$ for the comprehensibility subscale in the L1, L2 and the total groups} }
\label{tab_1}
\begin{center}
\begin{tabular}{|c|c|c|c|c|c|c|}
\hline
\hline
 {Group} &   {$N$} & {$ICC(A,1)$} & {$r_{WG}$} & {CV$\%$} & {$\widehat{d}$} &{$\widehat{d}^{*}$} \\
\cline{1-2}
\hline
\hline
L1	&  20 & 0.14& 0.90& 8.12& 0.17 & 0.19	\\
L2&   20 & 0.63& 0.84& 16.20 & 0.28& 0.32 \\
Total&  40 & 0.67& 0.87& 12.16& 0.22& 0.25\\
\hline
\hline
\end{tabular}
\end{center}
\end{table}
}




\section{Conclusions}
In this paper a measure of interrater absolute agreement for ordinal scales is proposed.
Such a  measure is not affected by restriction of variance problems and does not depend on the choice of a particular null
distribution. 
An unbiased estimator of the proposed measure is introduced and its sampling properties are investigated.
In the simulation study confidence intervals for the proposed interrater agreement index are
constructed using the normal approximation, the parametric and nonparametric bootstrap.
Furthermore, a pseudo-nonparametric bootstrap taking into account the sampling design is also implemented.
As previously stressed, the resampling involves both raters and targets sample.
Confidence intervals obtained with the normal approximation seem to perform very well both
in terms of coverage probability and computational cost.

{}


\begin{thebibliography}{}

\bibitem{booth:1994}
Booth, J. G., R. W. Butler, and P. Hall (1994). Bootstrap methods for finite
populations. \emph{Journal of the American Statistical Association}, 89 (428), 1282--1289. 

\bibitem{bove_2018} Bove, G., Nuzzo, E., Serafini, A. (2018)
Measurement of interrater agreement for the assessment of language proficiency. In: S. Capecchi, Di Iorio F., Simone R. \emph{ASMOD 2018: Proceedings of the Advanced Statistical Modelling for Ordinal Data Conference}. 
Universit\`a Federico II di Napoli, 24-26 October 2018. Napoli: FedOAPress, 61--68.



\bibitem{grilli:2002} Grilli L., Rampichini C. (2002) Scomposizione della dispersione per variabili statistiche
ordinali [Dispersion decomposition for ordinal variables], \emph{Statistica}, \textbf{62}, 111--116.


\bibitem{gross:1980} 
Gross, S. (1980). Median estimation in sample surveys. In \emph{Proceedings of the
Section on Survey Research Methods}, American Statistical Association, pp.
181--184.




\bibitem{demaree:1984}
James, L. J., Demaree, R. G.,Wolf, G. (1984). Estimating within-group interrater reliability with and without response bias.  \emph{Journal of Applied Psychology}, \textbf{69}, 85--98.

\bibitem{demaree:1993}
James L. J., Demaree R. G., Wolf G. (1993) rwg: An assessment of within-group interrater
agreement, \emph{Journal of Applied Psychology}, \textbf{78}, 306--309.

\bibitem{efron:1979}
Efron, B. (1979). Bootstrap methods: another look at the jackknife. \emph{The Annals
of Statistics}, \textbf{7}(1), 1--26.

\bibitem{kuiken:2017}
Kuiken F., Vedder I. (2017) Functional adequacy in L2 writing. Towards a new rating scale,
\emph{Language Testing}, \textbf{34}, 321-336.

\bibitem{Lebreton:2003}
LeBreton J.M., Burgess J.R.D., Kaiser R.B., Atchley E.K., James L.R. (2003) The restriction
of variance hypothesis and interrater reliability and agreement: Are ratings from
multiple sources really dissimilar?, \emph{Organizational Research Methods}, \textbf{6}, 80--128.

\bibitem{Lebreton:2008}
LeBreton J.M., Senter, J.L. (2008). Answers to 20 questions about interrater reliability and interrater agreement. \emph{Organizational Research Methods}, \textbf{11}(4), 815--852.


\bibitem{leti:1983}
Leti G. (1983) \emph{Statistica descrittiva}, Il Mulino, Bologna.


\bibitem{lomnicki:1952} 
Lomnicki Z.A. (1952) The Standard Error of Gini's Mean Difference. \emph{The Annals of Mathematical Statistics}, \textbf{23}, 14, 635--637.


\bibitem{mashreghi:2016}
Mashreghi, Z., Haziza, D., L\'eger, C. (2016).
A survey of bootstrap methods in finite population sampling. \emph{Statistics Surveys}, \textbf{10}, 1--52.

\bibitem{McGraw:1996}
McGraw K.O., Wong S.P. (1996) Forming inferences about some intraclass correlation coefficients,
\emph{Psychological Methods}, \textbf{1}, 30--46.

\bibitem{nuzzo:2018}
Nuzzo E., Bove G. (2018) Assessing functional adequacy across tasks: A comparison of
learners and native speakers' written texts, (submitted for publication).




\bibitem{piccarretta:2001}
Piccarretta, R. (2001). A new measure of nomila-ordinal association, \emph{Journal of Applied Statistics}, \textbf{28}, 1, 107--120. 


\bibitem{sheather:1991}
Sheather, S.J. and Jones, M.C. (1991). A Reliable Data-Based Bandwidth Selection Method
for Kernel Density Estimation. \emph{Journal of the Royal Statistical Society Series B}, \textbf{53},
683--690.

\bibitem{shrout:1979}
Shrout, P. E., and Fleiss, J. L. (1979). Intraclass correlations: Uses in assessing reliability. \emph{Psychological Bulletin}, \textbf{86}, 420--428.

\bibitem{von:2005}
von Eye A., Mun E.Y. (2005) \emph{Analyzing rater agreement. Manifest variable methods},
Lawrence Erlbaum Associates, Mahwah, New Jersey.

\end{thebibliography}
\end{document}